\documentclass[aps,prl,twocolumn,showpacs,superscriptaddress,letterpaper,preprintnumbers,amsmath,amssymb]{revtex4}  % for review and submission 
\usepackage{graphicx} 
\usepackage{dcolumn}  
\usepackage{bm}       
\usepackage{amssymb}  
\def\missET {{\not\!\! E_T}}
\begin{document}
\hspace{5.2in} \mbox{Fermilab-PUB-08-369-E}
\bibliographystyle{apsrev}

\title{Measurement of the top quark mass with dilepton events selected using neuroevolution at CDF}

\affiliation{Institute of Physics, Academia Sinica, Taipei, Taiwan 11529, Republic of China} 
\affiliation{Argonne National Laboratory, Argonne, Illinois 60439} 
\affiliation{University of Athens, 157 71 Athens, Greece} 
\affiliation{Institut de Fisica d'Altes Energies, Universitat Autonoma de Barcelona, E-08193, Bellaterra (Barcelona), Spain} 
\affiliation{Baylor University, Waco, Texas  76798} 
\affiliation{Istituto Nazionale di Fisica Nucleare Bologna, $^t$University of Bologna, I-40127 Bologna, Italy} 
\affiliation{Brandeis University, Waltham, Massachusetts 02254} 
\affiliation{University of California, Davis, Davis, California  95616} 
\affiliation{University of California, Los Angeles, Los Angeles, California  90024} 
\affiliation{University of California, San Diego, La Jolla, California  92093} 
\affiliation{University of California, Santa Barbara, Santa Barbara, California 93106} 
\affiliation{Instituto de Fisica de Cantabria, CSIC-University of Cantabria, 39005 Santander, Spain} 
\affiliation{Carnegie Mellon University, Pittsburgh, PA  15213} 
\affiliation{Enrico Fermi Institute, University of Chicago, Chicago, Illinois 60637} 
\affiliation{Comenius University, 842 48 Bratislava, Slovakia; Institute of Experimental Physics, 040 01 Kosice, Slovakia} 
\affiliation{Joint Institute for Nuclear Research, RU-141980 Dubna, Russia} 
\affiliation{Duke University, Durham, North Carolina  27708} 
\affiliation{Fermi National Accelerator Laboratory, Batavia, Illinois 60510} 
\affiliation{University of Florida, Gainesville, Florida  32611} 
\affiliation{Laboratori Nazionali di Frascati, Istituto Nazionale di Fisica Nucleare, I-00044 Frascati, Italy} 
\affiliation{University of Geneva, CH-1211 Geneva 4, Switzerland} 
\affiliation{Glasgow University, Glasgow G12 8QQ, United Kingdom} 
\affiliation{Harvard University, Cambridge, Massachusetts 02138} 
\affiliation{Division of High Energy Physics, Department of Physics, University of Helsinki and Helsinki Institute of Physics, FIN-00014, Helsinki, Finland} 
\affiliation{University of Illinois, Urbana, Illinois 61801} 
\affiliation{The Johns Hopkins University, Baltimore, Maryland 21218} 
\affiliation{Institut f\"{u}r Experimentelle Kernphysik, Universit\"{a}t Karlsruhe, 76128 Karlsruhe, Germany} 
\affiliation{Center for High Energy Physics: Kyungpook National University, Daegu 702-701, Korea; Seoul National University, Seoul 151-742, Korea; Sungkyunkwan University, Suwon 440-746, Korea; Korea Institute of Science and Technology Information, Daejeon, 305-806, Korea; Chonnam National University, Gwangju, 500-757, Korea} 
\affiliation{Ernest Orlando Lawrence Berkeley National Laboratory, Berkeley, California 94720} 
\affiliation{University of Liverpool, Liverpool L69 7ZE, United Kingdom} 
\affiliation{University College London, London WC1E 6BT, United Kingdom} 
\affiliation{Centro de Investigaciones Energeticas Medioambientales y Tecnologicas, E-28040 Madrid, Spain} 
\affiliation{Massachusetts Institute of Technology, Cambridge, Massachusetts  02139} 
\affiliation{Institute of Particle Physics: McGill University, Montr\'{e}al, Canada H3A~2T8; and University of Toronto, Toronto, Canada M5S~1A7} 
\affiliation{University of Michigan, Ann Arbor, Michigan 48109} 
\affiliation{Michigan State University, East Lansing, Michigan  48824}
\affiliation{Institution for Theoretical and Experimental Physics, ITEP, Moscow 117259, Russia} 
\affiliation{University of New Mexico, Albuquerque, New Mexico 87131} 
\affiliation{Northwestern University, Evanston, Illinois  60208} 
\affiliation{The Ohio State University, Columbus, Ohio  43210} 
\affiliation{Okayama University, Okayama 700-8530, Japan} 
\affiliation{Osaka City University, Osaka 588, Japan} 
\affiliation{University of Oxford, Oxford OX1 3RH, United Kingdom} 
\affiliation{Istituto Nazionale di Fisica Nucleare, Sezione di Padova-Trento, $^u$University of Padova, I-35131 Padova, Italy} 
\affiliation{LPNHE, Universite Pierre et Marie Curie/IN2P3-CNRS, UMR7585, Paris, F-75252 France} 
\affiliation{University of Pennsylvania, Philadelphia, Pennsylvania 19104} 
\affiliation{Istituto Nazionale di Fisica Nucleare Pisa, $^q$University of Pisa, $^r$University of Siena and $^s$Scuola Normale Superiore, I-56127 Pisa, Italy} 
\affiliation{University of Pittsburgh, Pittsburgh, Pennsylvania 15260} 
\affiliation{Purdue University, West Lafayette, Indiana 47907} 
\affiliation{University of Rochester, Rochester, New York 14627} 
\affiliation{The Rockefeller University, New York, New York 10021} 

\affiliation{Istituto Nazionale di Fisica Nucleare, Sezione di Roma 1, $^v$Sapienza Universit\`{a} di Roma, I-00185 Roma, Italy} 

\affiliation{Rutgers University, Piscataway, New Jersey 08855} 
\affiliation{Texas A\&M University, College Station, Texas 77843} 
\affiliation{Istituto Nazionale di Fisica Nucleare Trieste/\ Udine, $^w$University of Trieste/\ Udine, Italy} 
\affiliation{University of Tsukuba, Tsukuba, Ibaraki 305, Japan} 
\affiliation{Tufts University, Medford, Massachusetts 02155} 
\affiliation{Waseda University, Tokyo 169, Japan} 
\affiliation{Wayne State University, Detroit, Michigan  48201} 
\affiliation{University of Wisconsin, Madison, Wisconsin 53706} 
\affiliation{Yale University, New Haven, Connecticut 06520} 
\author{T.~Aaltonen}
\affiliation{Division of High Energy Physics, Department of Physics, University of Helsinki and Helsinki Institute of Physics, FIN-00014, Helsinki, Finland}
\author{J.~Adelman}
\affiliation{Enrico Fermi Institute, University of Chicago, Chicago, Illinois 60637}
\author{T.~Akimoto}
\affiliation{University of Tsukuba, Tsukuba, Ibaraki 305, Japan}
\author{M.G.~Albrow}
\affiliation{Fermi National Accelerator Laboratory, Batavia, Illinois 60510}
\author{B.~\'{A}lvarez~Gonz\'{a}lez}
\affiliation{Instituto de Fisica de Cantabria, CSIC-University of Cantabria, 39005 Santander, Spain}
\author{S.~Amerio$^u$}
\affiliation{Istituto Nazionale di Fisica Nucleare, Sezione di Padova-Trento, $^u$University of Padova, I-35131 Padova, Italy} 

\author{D.~Amidei}
\affiliation{University of Michigan, Ann Arbor, Michigan 48109}
\author{A.~Anastassov}
\affiliation{Northwestern University, Evanston, Illinois  60208}
\author{A.~Annovi}
\affiliation{Laboratori Nazionali di Frascati, Istituto Nazionale di Fisica Nucleare, I-00044 Frascati, Italy}
\author{J.~Antos}
\affiliation{Comenius University, 842 48 Bratislava, Slovakia; Institute of Experimental Physics, 040 01 Kosice, Slovakia}
% ...

\collaboration{CDF Collaboration\footnote{With visitors from $^a$Universiteit Antwerpen, B-2610 Antwerp, Belgium, 
$^b$Chinese Academy of Sciences, Beijing 100864, China, 
$^c$University of Bristol, Bristol BS8 1TL, United Kingdom, 
$^d$University of California Irvine, Irvine, CA  92697, 
$^e$University of California Santa Cruz, Santa Cruz, CA  95064, 
$^f$Cornell University, Ithaca, NY  14853, 
$^g$University of Cyprus, Nicosia CY-1678, Cyprus, 
$^h$University College Dublin, Dublin 4, Ireland, 
$^i$University of Edinburgh, Edinburgh EH9 3JZ, United Kingdom, 
$^j$Universidad Iberoamericana, Mexico D.F., Mexico, 
$^{jk}$Informatics Institute, University of Amsterdam, Amsterdam, The Netherlands,
$^k$University of Manchester, Manchester M13 9PL, England, 
$^l$Nagasaki Institute of Applied Science, Nagasaki, Japan, 
$^m$University de Oviedo, E-33007 Oviedo, Spain, 
$^n$Queen Mary, University of London, London, E1 4NS, England, 
$^o$Texas Tech University, Lubbock, TX  79409, 
$^p$IFIC(CSIC-Universitat de Valencia), 46071 Valencia, Spain,
$^x$Royal Society of Edinburgh/Scottish Executive Support Research Fellow,
$^y$Istituto Nazionale di Fisica Nucleare, Sezione di Cagliari, 09042 Monserrato (Cagliari), Italy 
}}
\noaffiliation

\begin{abstract}
We report a measurement of the top quark mass $M_t$ in the dilepton decay channel $t\bar{t}\rightarrow b\ell'^{+}\nu'_\ell\overline{b}\ell^{-}\overline{\nu}_{\ell}$. Events are selected with a neural network which has been directly optimized for statistical precision in top quark mass using neuroevolution, a technique modeled on biological evolution.  The top quark mass is extracted from per-event probability densities that are formed by the convolution of leading order matrix elements and detector resolution functions. The joint probability is the product of the probability densities from 344 candidate events in 2.0~fb$^{-1}$ of $p\bar{p}$ collisions collected with the CDF II detector, yielding a measurement of $M_t= 171.2\pm 2.7(\textrm{stat.})\pm 2.9(\textrm{syst.})\mathrm{~GeV}/c^2$.
\end{abstract}


%for revtex
\maketitle 

 Over ten years after the discovery of the top quark, its mass, $M_t$, remains a quantity of great interest.  $M_t$-dependent terms contribute to radiative corrections to precision electroweak observables, thus providing information on the unobserved Higgs boson~\cite{higgsconstr} and other particles in possible extensions to the standard model~\cite{topsusy} (SM). Top quarks are produced only at the Fermilab Tevatron, primarily in pairs and decay $\approx 100\%$ to a $W$ boson and a $b$ quark, $t\bar{t}\to W^{+}bW^{-}\bar{b}$, in the SM. The dilepton channel, where both $W$ bosons decay to charged leptons (electrons and muons, including leptonic decays of $\tau$ leptons) and neutrinos, has the smallest branching fraction, but also has the least number of hadronic jets in the final state and hence a smaller sensitivity to their energy calibration. Significant differences in the measurements of $M_t$ in different decay channels could indicate contributions from sources beyond the SM~\cite{newsources}.

Reconstruction of $M_t$ in the dilepton channel presents unique challenges, as the two neutrinos in the final state result in a kinematically underconstrained system. We utilize a likelihood-based estimator that convolutes leading order SM matrix elements and detector resolution functions and integrates over unmeasured quantities. Prior applications of this method to dilepton events have yielded the most precise measurements of $M_t$ in this channel~\cite{Abulencia:2005uq,Abulencia:2006qj,prd1fb}. These prior measurements utilize  event selection criteria that were designed to maximize signal purity for a measurement of the $t\bar{t}$ production cross section~\cite{Acosta:2004uw}.  The selection optimization for precision in $M_t$ is hampered by the difficulty of searching the space of arbitrary multivariate selections. Well established multivariate algorithms such as neural networks are typically limited to minimization of a specific metric, such as misclassification error. They are not designed to optimize an event ensemble property, such as the uncertainty on the top quark mass. In contrast, the technique of neuroevolution~\cite{neuroevolution} combines the parametrization of an abitrary multivariate selection described by a neural network with an evolutionary minimization approach to search for the network weights and topology which optimizes an arbitrary metric.    In this Letter, we present a measurement using an improved matrix element analysis technique and an event selection optimized with neuroevolution to minimize the expected statistical uncertainty in the top quark mass measurement. %This novel application of neuroevolution yields an event selection with markedly poorer signal %purity, yet significantly smaller expected uncertainty in $M_t$. 
We utilize 2.0~fb$^{-1}$ of data collected between March 2002 and May 2007 with the CDF II detector at the Fermilab Tevatron.

CDF II~\cite{Acosta:2004yw,wzxsection,coordinates} contains a charged particle tracking system consisting of a silicon microstrip tracker and a drift chamber immersed in a 1.4~T magnetic field. Surrounding electromagnetic and hadronic calorimeters measure particle energies. Outside the calorimeters, drift chambers and scintillators detect muons. 

We use lepton triggers that require an electron or muon with $p_T>18$~GeV$/c$. We define a preselection which satisfies the basic signature of top dilepton decay: two oppositely charged leptons with $p_T>20$~GeV$/c$, two or more jets with $E_T>15$~GeV~\cite{jetclu} within the region $|\eta|<2.5$, $\missET>20$~GeV~\cite{metdef}, and dilepton invariant mass $M_{ll}>10$~GeV/$c^2$. Suppression of the $Z\rightarrow ll$ background is performed by the subsequent neural-network selection. 

\begin{figure}[t]
\begin{center}
\includegraphics[width=0.48\columnwidth,angle=90]{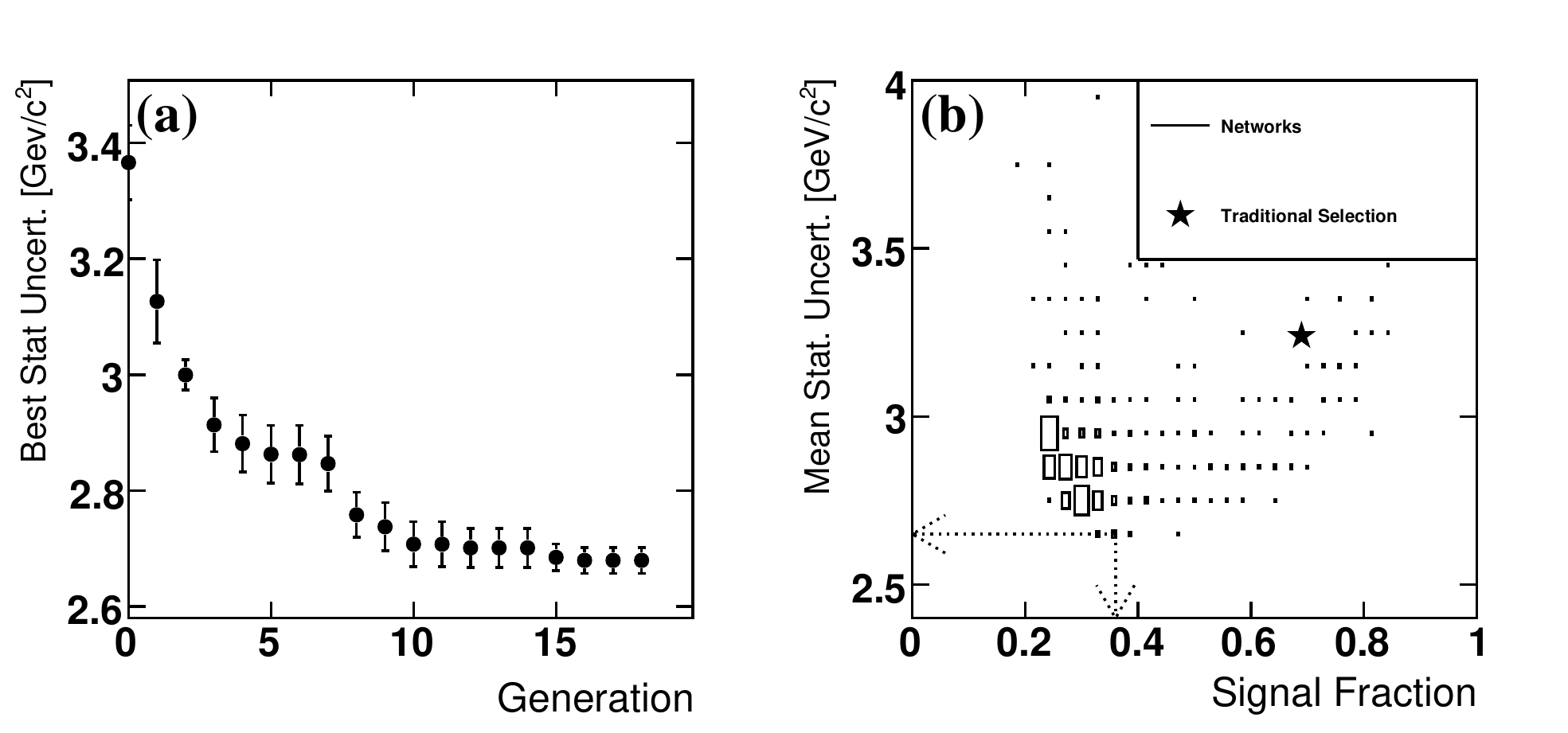}
\caption{ Top,  expected statistical uncertainty for the best network in each successive generation of network evaluation. The points show the average performance for each generation; the error bars show the variation due to the randomly generated networks in generation 0. 
Bottom, expected statistical uncertainty on $M_{t}$ versus signal fraction after neural network selection, for all evaluated networks. The selection~\cite{Acosta:2004uw} used in previous measurements is shown ($\star$) for comparison. The arrows show the expected statistical uncertainty and signal fraction corresponding to the network used in the
 analysis. }

\label{fig:evolution}
\end{center}
\end{figure}

Neuroevolution, an approach modeled on biological evolution, is used to search directly for the optimal neural network.   Beginning with a population of 150 networks with random weights, the statistical precision of $M_t$ is evaluated for each network by performing experiments using the simulated signal and background events which survive a threshold requirement on the network output. The events are simulated using the {\sc pythia}~\cite{pythia} and {\sc alpgen}~\cite{alpgen} generators and a full detector simulation~\cite{cdfsim}. Poorly performing networks are culled and the 30 strongest performers are bred together and mutated in successive generations until performance reaches a plateau in a statistically independent pool of events, which occurs after 15 generations. The statistical uncertainty obtained from the best performer in each generation is shown in Fig.~\ref{fig:evolution}(a). In the context of an arbitrary but {\it a priori} fixed choice of network threshold, the networks evolve to optimize the selection regardless of the threshold's value. Because we have optimized directly on the final statistical precision rather than some intermediate or approximate figure of merit, the best-performing network is the one which gives the most precise measurement. This approach has been shown to significantly outperform traditional methods in event selection~\cite{DBLP:conf/aaai/2007}. In particular, we use neuroevolution of augmenting topologies (NEAT)~\cite{neat}, a neuroevolutionary method capable of evolving a network's topology and weights. 

\begin{figure}[t]
\begin{center}
\includegraphics[width=0.9\columnwidth]{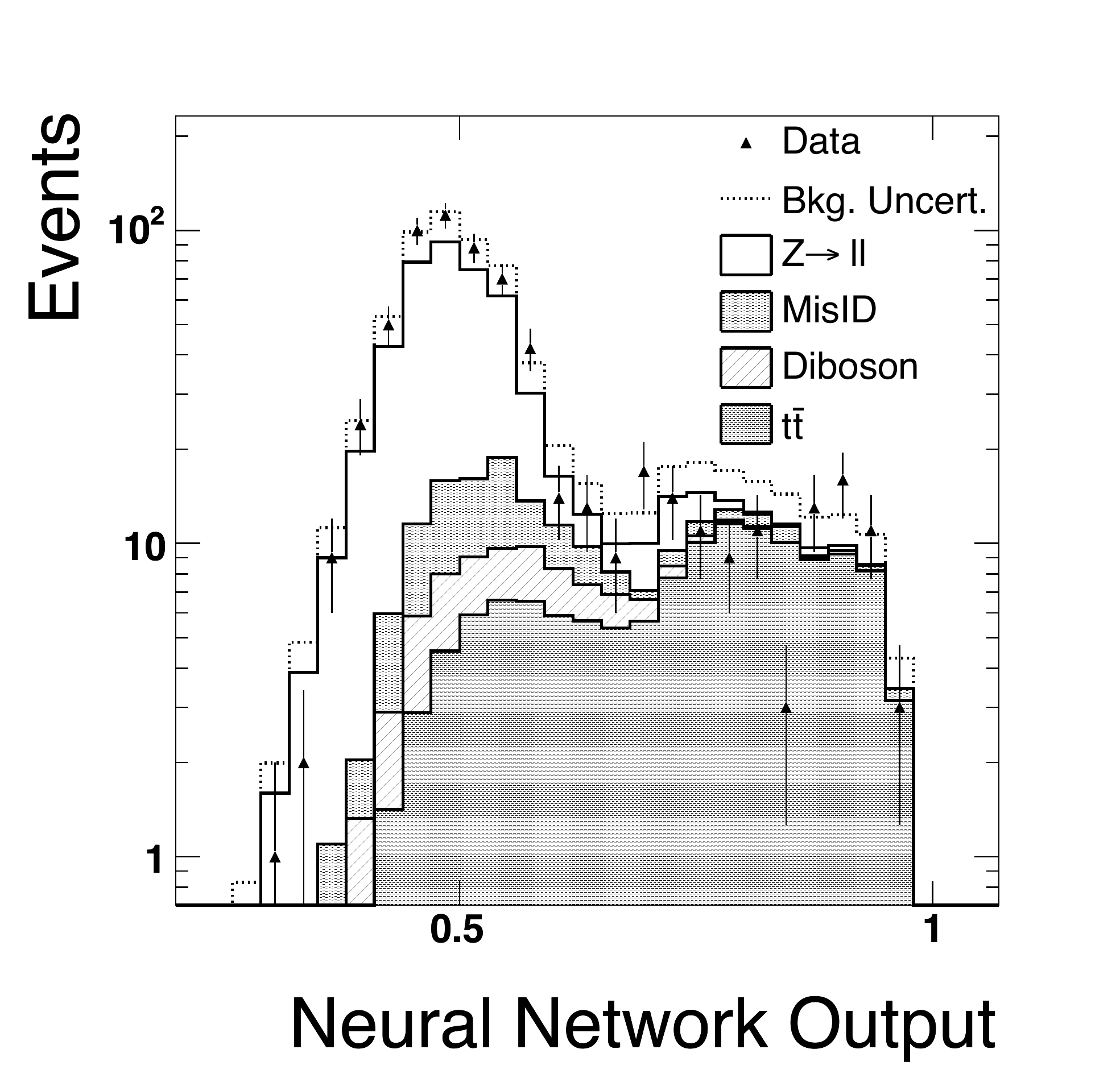}
\caption{The output of the final network evaluated on the
collected data (black triangles), with expected signal and background contributions (stacked solid histograms). The data show events passing the pre-selection. The evolution of the optimum selection network is performed with an {\it a priori} threshold set at 0.5 for candidate selection. Of the 642 pre-selected events shown, 344 events pass this threshold and constitute the final candidate sample for mass-fitting.}
\label{fig:data}
\end{center}
\end{figure}

Some of the events passing this selection have secondary vertex tags~\cite{acosta:052003}, which enhance b-quark fraction and thus signal purity. We exploit this enhancement by
 separately fitting events with and without secondary vertex tags, and combining the fits. 
The predicted number of signal and background events is shown in Table~\ref{tab:samplecomp}. Using the optimized selection improves the {\it a priori} statistical uncertainty on $M_t$ over the selection used in previous analyses~\cite{prd1fb} by 20\%.   This neural network selection yields 344 candidate events (Fig.~\ref{fig:data}).  Strikingly, the sample selected by the neural network is expected to be dominated by background events; the resulting measurement is expected to be more precise than previous measurements due to the increase in $t\overline{t}$ acceptance and the suppression of background effects as described below. The distribution of expected statistical uncertainty versus signal purity for all evaluated networks can be seen in Fig.~\ref{fig:evolution}(b).

\begin{table}[t]
\caption{Expected sample composition after neural network selection for events with and without secondary vertex tags.}
\begin{ruledtabular}
\begin{tabular}{lrr}
Source             & $N$(0-tag)   & $N(\ge$ 1-tag)       \\ \hline
$ Z\rightarrow ll $     & $116.5\pm18.6$ & $4.1\pm1.8$   \\ 
$ Z\rightarrow ll +c\bar{c}/b\bar{b}$  & $9.3\pm1.4$   & $10.1\pm4.0$   \\ 
$ WW,WZ,ZZ,W\gamma$   & $17.3\pm5.9$  & $0.7\pm0.7$   \\ 
 Misidentified leptons       & $29.0\pm8.7$  & $4.5\pm1.1$   \\ 
$ t\bar{t}$ ($\sigma=6.7$~pb, $M_t=175$~GeV$/c^2$)   & $43.8\pm4.4$  & $78.0\pm6.2$  \\ 
\hline
Total                   & $215.8\pm21.9$        & $97.5\pm7.2$  \\
Observed (2.0 fb$^{-1}$)            & 246   & 98    
\end{tabular}
\end{ruledtabular}
\label{tab:samplecomp}
\end{table}

We express the probability density for the observed lepton and jet measurements, $\mathbf{x}_i$, as a function of the top quark mass $M_t$ as $P_s(\mathbf{x}_i|M_t)$.  We calculate $P_s(\mathbf{x}_i|M_t)$ using the theoretical description of the $t\bar{t}$ production and decay process with respect to $\mathbf{x}_i$, $P_s(\mathbf{x}_i | M_t) = [1/\sigma(M_t)] [d\sigma(M_t)/d\mathbf{x}_i] $, where $\frac{ d\sigma}{d\mathbf{x}_i}$ is the differential cross section~\cite{meref1, meref2, glueglue} and $\sigma$ is the total cross section. The term $1/\sigma(M_t)$ ensures that the probability density satisfies the normalization condition, $\int d\mathbf{x}_i\ P_s(\mathbf{x}_i|M_t) = 1$. 

We evaluate $P_s(\mathbf{x}_i|M_t)$~\cite{prd1fb} by integrating over quantities that are not directly measured, such as neutrino momenta and quark energies. The effect of simplifying assumptions is estimated using simulated experiments. We integrate over quark energies using a parameterized detector transfer function~\cite{Abulencia:2006qj} $W(p,j)$, defined as the probability of measuring jet energy $j$ given quark energy $p$. 

We account for backgrounds using their probability densities $P_{bg_k}(\mathbf{x}_i)$ and form the full per-event probability
\begin{equation}
P^{n}(\mathbf{x}_i|M_t) = P_{s}(\mathbf{x}_i|M_t) p^{n}_s + \sum_k P_{bg_k}(\mathbf{x}_i) p^{n}_{bg_k}. 
\label{eqn:combination}
\end{equation}

\noindent The functions $P_{bg_k}(\mathbf{x}_i)$ are calculated using the  differential cross-section for each background.  The proportions  $p^{n}_s$ and $p^{n}_{bg_k}$ depend on whether the event has $n$ secondary vertex tags, and are obtained from Table~\ref{tab:samplecomp}.
We evaluate background probability densities for: $Z/\gamma^* (\to ee,\mu\mu) +$jets, $W + \geq 3$ jets where a jet is misidentified as a lepton, and $WW +$jets. Probability densities for smaller backgrounds ($WZ$, $ZZ$, $W\gamma$, and $Z\rightarrow\tau\tau$) provide negligible gain in sensitivity and are not modeled.

The posterior joint probability for the sample is the product of the per-event probability densities,
\begin{equation}
P(\mathbf{x} | M_t) = \Big[\prod_{i_0}P^{0}(\mathbf{x}_{i_0} | M_t)\Big]\times  \Big[\prod_{i_1}P^{\geq 1}(\mathbf{x}_{i_1} | M_t)\Big] 
\end{equation}
over all untagged ($i_0$) and tagged ($i_1$) events. The measured mass $M_t$ is taken as the mean $\langle M_t\rangle$ computed using the posterior probability, and the measured statistical uncertainty $\Delta M_t$ is taken as the standard deviation.

\begin{figure}[!ht]
\begin{center}
\includegraphics[width=0.49\columnwidth]{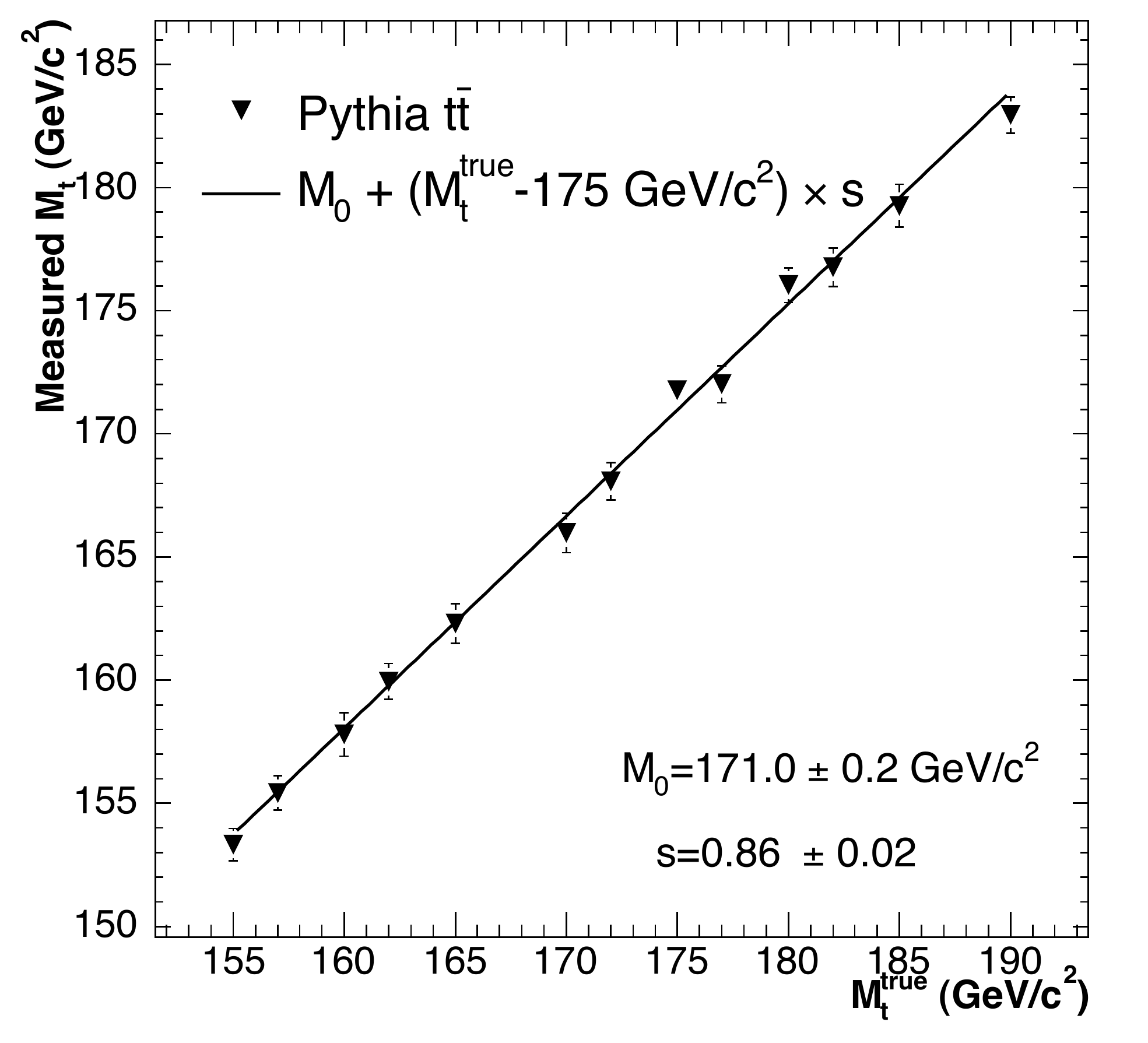}
\put(-110,-10){(a)}
\includegraphics[width=0.49\columnwidth]{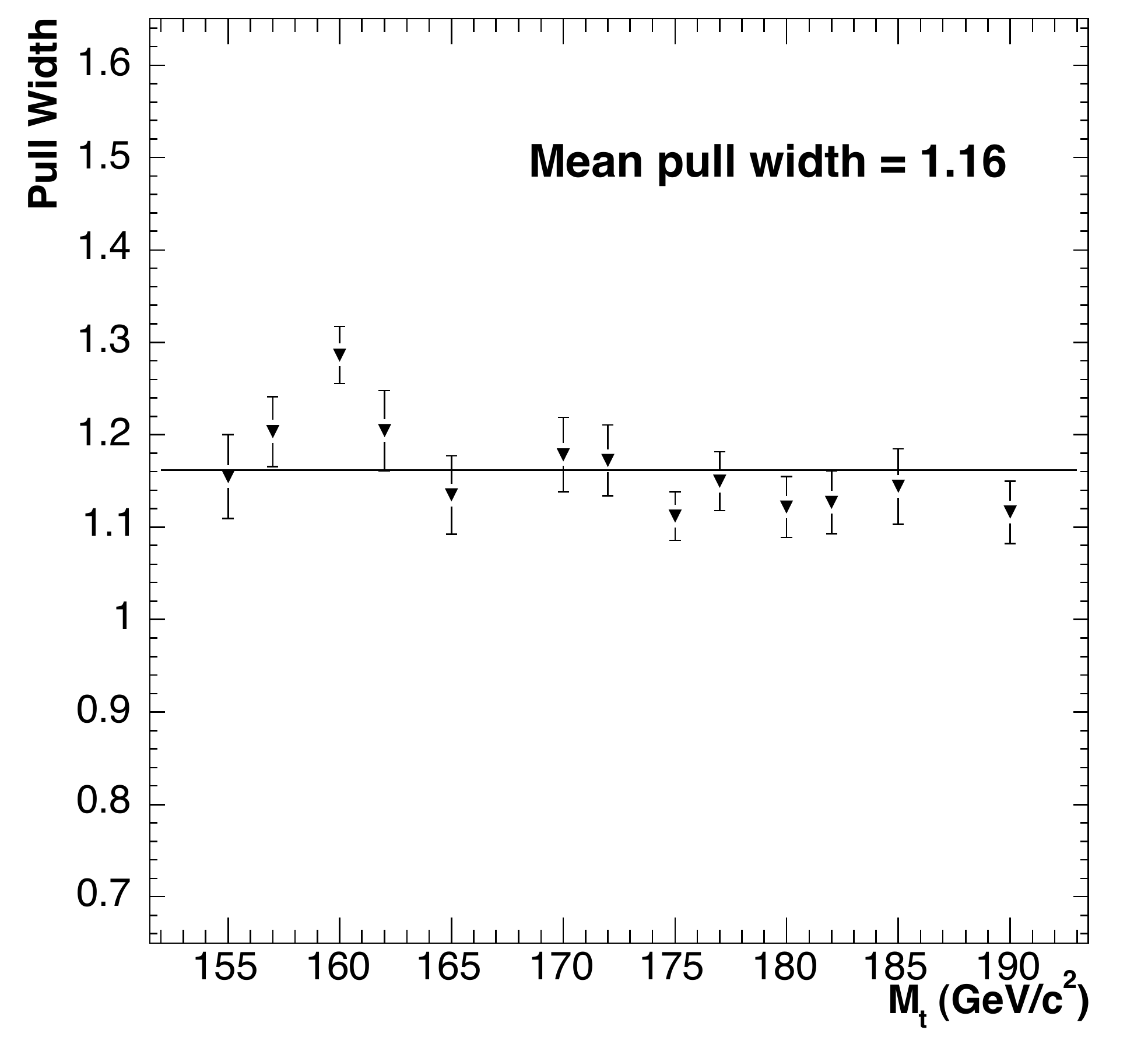}
\put(-110,-10){(b)}
\caption{{\it (a)} Mean measured $M_t$ in simulated experiments versus top quark masses. The solid line is a linear fit to the points. {\it (b)} Pull widths from simulated experiments versus top quark masses. The solid line is the average over all points.}
\label{fig:response}
\end{center}
\end{figure}

The response of our method for simulated experiments (Fig.~\ref{fig:response}a) is consistent with a linear dependence on the true top mass. Its slope is less than unity  due to the presence of unmodeled background. We derive corrections, $M_t\to 175.0\textrm{ GeV}/c^2+(M_t-171.0\textrm{ GeV}/c^2)/0.86$ and $\Delta M_t\to \Delta M_t/0.86$, from this response and apply them to the measured quantities in data.

From the pull distribution of our simulated experiments, we find that $\Delta M_t$ is underestimated (Fig.~\ref{fig:response}b). This is due to simplifying assumptions made in the probability calculations for computational tractability~\cite{Abulencia:2006qj}. These assumptions are violated in small, well-understood ways in realistic events. We scale $\Delta M_t$ by an additional factor, $S=1.16$, derived from our simulated experiments.
Applying this method   to the 344 candidate events, we measure $M_{t}=171.2\pm 2.7 (\textrm{stat.})\textrm{ GeV}/c^2$. The posterior probability is Gaussian within the statistical accuracy of the Monte
Carlo integration.

\begin{table}[b]
\caption{Summary of systematic uncertainties on the measured top quark mass.}

\begin{ruledtabular}

\begin{tabular}{lr}

Source & Size (GeV/$c^2$)\\\hline
Generic jet energy scale & 2.5\\
$b$-Jet Energy Scale & 0.4 \\
In-time pileup & 0.2 \\
Generator & 0.9\\
PDFs & 0.6\\
Background statistics & 0.5\\
Radiation & 0.5\\
Response correction & 0.4\\
Sample composition uncertainty & 0.3 \\
Background modeling & 0.2\\
Lepton energy scale & 0.1\\\hline
Total &  2.9\\

\end{tabular}
\end{ruledtabular}
\label{tab:sys} \rm
\end{table}

There are several sources of systematic uncertainty in our measurement, which are summarized in Table~\ref{tab:sys}. The single largest source of systematic error comes from the uncertainty in the jet energy scale, which we estimate to be 2.5~GeV$/c^2$ by varying the scale within its uncertainty~\cite{jes}. An uncertainty specific to jets resulting from $b$ partons contributes 0.4~GeV$/c^2$ while in-time pileup contributes 0.2~GeV$/c^2$.  Uncertainty due to the Monte Carlo generator used for $t\bar{t}$ events is estimated as the difference in $M_t$ extracted from {\sc pythia} events and {\sc herwig}~\cite{herwig} events and amounts to 0.9~GeV$/c^2$.  Uncertainties due to PDFs are estimated using different PDF sets ({\sc cteq5l}~\cite{cteq} vs. {\sc mrst72}~\cite{mrst}), different values of $\Lambda_{QCD}$, and varying the eigenvectors of the {\sc cteq6m}~\cite{cteq} set; the quadrature sum of the latter two (dominant) uncertainties is 0.6~GeV$/c^2$. The limited number of background events available for simulated experiments results in an uncertainty on the shape of the background distributions, which yields an uncertainty on $M_{t}$ of 0.5~GeV$/c^2$. Uncertainty due to imperfect modeling of initial and final state  QCD radiation (ISR and FSR, respectively) is estimated by varying the amounts of ISR and FSR in simulated events~\cite{ljetsprd} and is estimated to be 0.5~GeV$/c^2$. The uncertainty in the mass due to uncertainties in the response correction is evaluated by varying the response within the uncertainties shown in Fig.~\ref{fig:response}a and is 0.4~GeV$/c^2$.  The contribution from uncertainties in background composition is estimated by varying the background normalizations from Table~\ref{tab:samplecomp} within their uncertainties and amounts to 0.3 GeV$/c^2$. We estimate the uncertainty coming from  modeling of the missing tranverse energy in $Z/\gamma^*$ events and the uncertainty in the data-derived model of misindentified leptons to be 0.2~GeV$/c^2$. The uncertainty in the lepton energy scale contributes an uncertainty of 0.1~GeV$/c^2$ to our measurement. Adding in quadrature yields a total systematic uncertainty of 2.9 GeV/$c^2$.

In summary, we have presented a new measurement of the top quark mass in the dilepton channel. We have applied the technique of neuroevolution, for the first time in particle physics, to devise an event selection criterion which optimizes statistical precision. We measure $M_t= 171.2\pm 2.7(\textrm{stat.})\pm 2.9(\textrm{syst.})\mathrm{~GeV}/c^2$. This is the single most precise measurement of $M_t$ in this channel to date, is in good agreement with measurements in other channels~\cite{aaltonen:182002,aaltonen:072009}, and represents a $\sim$30\% improvement in statistical precision over the previously published measurements in this channel~\cite{prd1fb,tuula, d0dilep2}. 

We thank the Fermilab staff and the technical staffs of the participating institutions for their vital contributions. This work was supported by the U.S. Department of Energy and National Science Foundation; the Italian Istituto Nazionale di Fisica Nucleare; the Ministry of Education, Culture, Sports, Science and Technology of Japan; the Natural Sciences and Engineering Research Council of Canada; the National Science Council of the Republic of China; the Swiss National Science Foundation; the A.P. Sloan Foundation; the Bundesministerium f\"ur Bildung und Forschung, Germany; the Korean Science and Engineering Foundation and the Korean Research Foundation; the Science and Technology Facilities Council and the Royal Society, UK; the Institut National de Physique Nucleaire et Physique des Particules/CNRS; the Russian Foundation for Basic Research; the Ministerio de Educaci\'{o}n y Ciencia and Programa Consolider-Ingenio 2010, Spain; the Slovak R\&D Agency; and the Academy of Finland.

\bibliography{prl}

\begin{thebibliography}{31}
\bibitem[{coo()}]{coordinates}
\bibinfo{note}{CDF uses a cylindrical coordinate system with the $z$ axis along
  the proton beam axis. Pseudorapidity is $\eta \equiv -\ln(\tan \theta/2)$,
  $\theta$ ($\phi$) is the polar (azimuthal) angle relative to the $z$ axis,
  while $p_T = |p|\ \sin \theta$, $E_T = E\ \sin \theta$.}
\bibitem[{met()}]{metdef}
\bibinfo{note}{Missing transverse energy $\missET \equiv | -\sum_i E_T^i \vec
  n_i |$, where $E_T^i$ is the transverse energy in calorimeter tower $i$, and
  $\vec n_i$ is the unit transverse vector from the beamline to tower $i$.}
\bibitem[{glu()}]{glueglue}
\bibinfo{note}{While $\approx$15\% of $t\overline{t}$ are produced by
  gluon-gluon fusion ($gg\to t\overline{t}$), it has negligible effect on our
  measurement.}
\end{thebibliography}

\end{document}